\documentclass[fleqn,10pt]{wlscirep}
\usepackage[utf8]{inputenc}
\usepackage[T1]{fontenc}
\usepackage{tikz}
\usepackage{amsmath,amssymb,tabularx}
\usepackage{graphicx}
\usepackage{braket}
\usepackage{float}

\title{GEOM: Energy-annotated molecular conformations for property prediction and molecular generation}

\author[1,2]{Simon Axelrod}
\author[2]{Rafael G\'{o}mez-Bombarelli}
\affil[1]{Harvard University, Department of Chemistry and Chemical Biology, Cambridge, MA, 02138, USA}
\affil[2]{Massachusetts Institute of Technology, Department of Materials Science and Engineering, Cambridge, MA, 02139, USA}

\affil[*]{corresponding author: Rafael G\'{o}mez-Bombarelli (rafagb@mit.edu)}

\begin{abstract}
Machine learning (ML) outperforms traditional approaches in many molecular design tasks. ML models usually predict molecular properties from a 2D chemical graph or a single 3D structure, but neither of these representations accounts for the ensemble of 3D conformers that are accessible to a molecule. Property prediction could be improved by using conformer ensembles as input, but there is no large-scale dataset that contains graphs annotated with accurate conformers and experimental data. Here we use advanced sampling and semi-empirical density functional theory (DFT) to generate 37 million molecular conformations for over 450,000 molecules. The Geometric Ensemble Of Molecules (GEOM) dataset contains conformers for 133,000 species from QM9, and 317,000 species with experimental data related to biophysics, physiology, and physical chemistry. Ensembles of 1,511 species with BACE-1 inhibition data are also labeled with high-quality DFT free energies in an implicit water solvent, and 534 ensembles are further optimized with DFT. GEOM will assist in the development of models that predict properties from conformer ensembles, and generative models that sample 3D conformations.
\end{abstract}
\begin{document}

\flushbottom
\maketitle

\thispagestyle{empty}

\section*{Background \& Summary}

Accurate and affordable prediction of molecular properties is a longstanding goal of computational chemistry. Predictions can be generated with rule-based \cite{norinder2006discrimination} or physics-based \cite{durrant2011molecular} methods, which typically involve a trade-off between accuracy and speed. Machine learning offers an attractive alternative, as it is far quicker than physics-based methods and outperforms traditional rule-based baselines in many molecule-related tasks, including property prediction and virtual screening \cite{ecoli_1,Gomez-Bombarelli2016,Zhavoronkov2019}, inverse design using generative models  \cite{Schwalbe-Koda2019GenerativeDesign,Gomez-Bombarelli2018,Jin2018,DeCao2018,Li2018LearningGraphs,Dai2018,Wang2020,Noe2019}, reinforcement learning \cite{Olivecrona2017a,Gottipati2020,Guimaraes2017,Popova2018}, differentiable simulators \cite{Wang2020,AlQuraishi2019,JohnIngraham019}, and synthesis planning and retrosynthesis \cite{Segler2018Planning,Coley2017a}. 

\begin{figure}[H]
    \centering
    \includegraphics[width=0.7\textwidth]{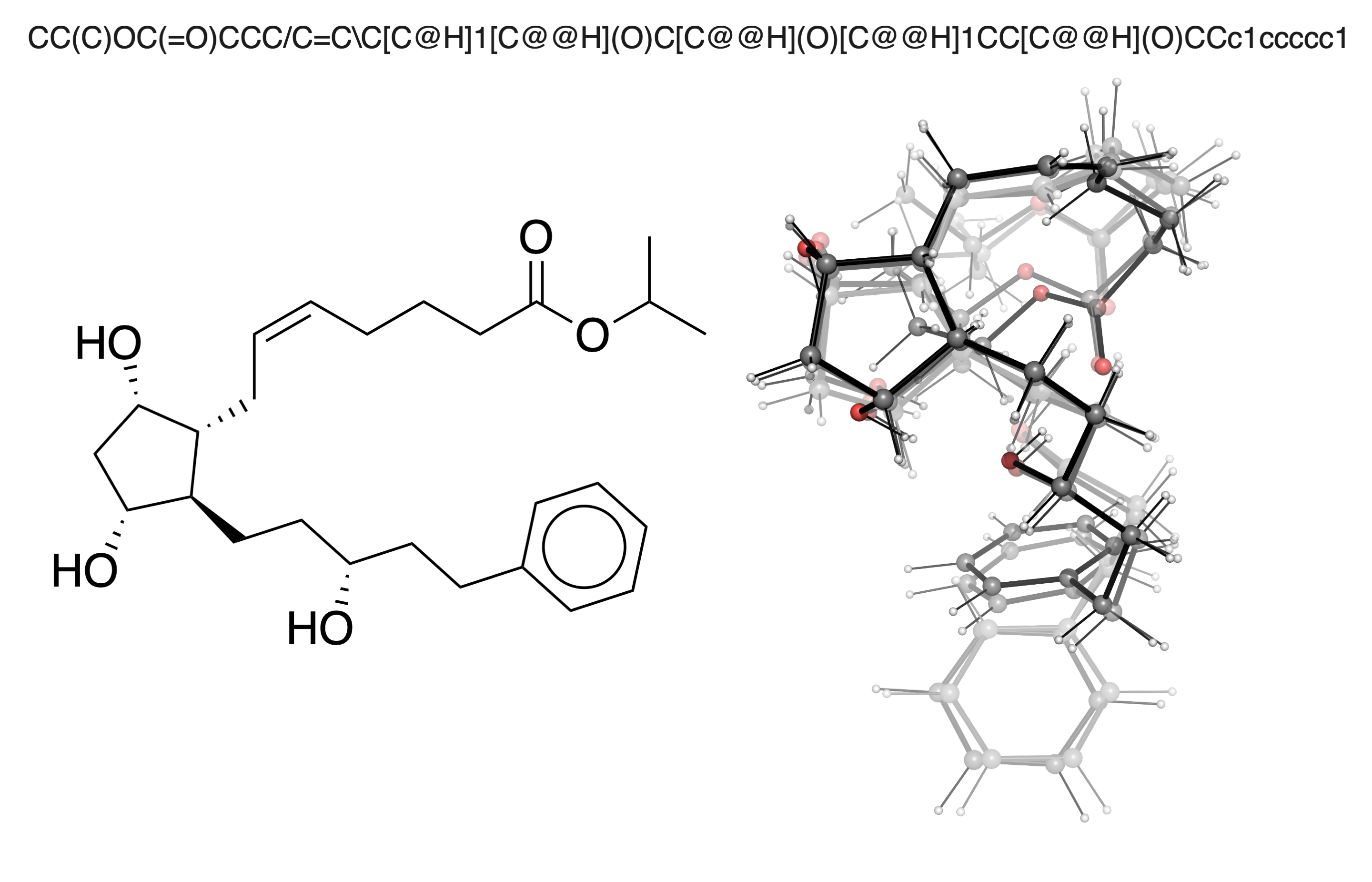}
    \caption{Molecular representations of the latanoprost molecule. \textit{top} SMILES string. \textit{left} Stereochemical formula with edge features, including wedges for in- and out-of-plane bonds, and a double line for \textit{cis} isomerism. \textit{right} Overlay of conformers. Higher transparency corresponds to lower statistical weight. }
    \label{fig:molecularrepresentations}
\end{figure}

Advances in molecular machine learning have been enabled by algorithmic improvements \cite{duvenaud_convolutional_2015, Kearnes2016, Yang2019Are, Anderson2019, Thomas2018, dimenet} and by reference datasets and tasks \cite{Ramsundar-et-al-2019}. A number of reference datasets provide unlabeled molecules for generation tasks \cite{chembl2018, Sterling2015, Gomez-Bombarelli2018, Brown2019GuacaMol, Polykovskiy2018Molecular} or experimentally labeled molecules for property prediction \cite{Delaney2004, Mobley2014, Wang2004, Wu2018MoleculeNet, Ramsundar-et-al-2019}. The molecules are typically represented as SMILES \cite{Weininger1988} or InChi \cite{Heller2015} strings, which can be converted into 2D graphs, or as single 3D structures. These representations can be used as input to machine learning models that predict properties or generate new compounds. However, these representations fail to capture the flexibility of molecules, which consist of atoms in continual motion on a potential energy surface (PES). Molecular properties are a function of the conformers accessible at finite temperature \cite{kuhn2016real, hawkins2017conformation}, which are not explicitly included in a 2D or single 3D representation (Fig. \ref{fig:molecularrepresentations}). Models that map conformer ensembles to experimental properties could be of interest, but they require a dataset with both conformers and experimental data.

Here we present the Geometric Ensemble Of Molecules (GEOM), a dataset of high-quality conformers for 317,928 mid-sized organic molecules with experimental data, and 133,258 molecules from the QM9 dataset \cite{Ramakrishnan2014}. 304,466 drug-like species and their biological assay results were accessed as part of AICures \cite{aicures}, an open machine learning challenge to predict which drugs can be repurposed to treat COVID-19 and related illnesses. 16,865 molecules are from the MoleculeNet benchmark \cite{Wu2018MoleculeNet}. They are labeled with experimental properties related to physical chemistry, biophysics, and physiology. Conformers were generated with the CREST program \cite{CREST}, which uses extensive sampling based on the semi-empirical extfended tight-binding method (GFN2-xTB \cite{bannwarth2019gfn2}) to generate reliable and accurate structures. CREST ensembles from 1,511 species in the BACE dataset \cite{subramanian2016computational} were also labeled with high-accuracy single-point DFT energies and semi-empirical quasi-harmonic free energies. Of these ensembles, 534 were further refined with DFT geometry optimizations.

GEOM addresses two key gaps in the dataset literature. First, the data can be used to benchmark new models that take conformers as input to predict experimental properties, such as biological assay results for antiviral activity, or physicochemical and physiological properties. Such models could not be trained on the above molecular datasets, which contain only 2D graphs or single 3D structures. Some datasets provide single 3D structures for hundreds of thousands of molecules \cite{Ramakrishnan2014, gravzulis2009crystallography, groom2016cambridge}, but do not include a full ensemble for each species. Others contain a continuum of high-quality 3D structures for each species, but only contain hundreds of molecules \cite{Smith2017,Smith2017b,Smith2018, Chmiela2017, simm2019generative, Kanal2018}. Yet others contain conformers for tens of thousands of molecules with experimental data \cite{bolton2011pubchem3d}, but the conformers are of force-field quality (see below). GEOM is unique in its size, number of conformers per species, conformer quality, and connection with experiment.

Second, GEOM can be used to train generative models to predict conformers given an input molecular graph. This is an active area of research that seeks to lower the computation cost compared to exhaustive torsional approaches and to increase the speed, reliability and accuracy compared to stochastic approaches \cite{Hoffmann2019,Simm2020,Stieffenhofer2020,Hoffmann2019a,Imrie2020,Mansimov2019,Chan2019,Gebauer2019Symmetry,Gebauer2018Generating,Wang2018Coarse}. The size and simulation accuracy of the GEOM dataset make it an ideal training set and for pre-training generalizable models. Moreover, machine learning models for conformer generation are orders of magnitude faster than the methods used to generate GEOM. Hence models trained on GEOM may be able to reproduce its accuracy on unseen molecules at a fraction of the cost. As discussed below, the CREST ensembles have high coverage of the true thermally accessible conformers. Hence GEOM is an excellent benchmark for the recall and diversity of conformer generation methods. However, the CREST statistical weights for each conformer are rather inaccurate. Therefore, benchmarks that include conformer probabilities should use the DFT weights provided in GEOM.

Table \ref{tab:descriptor_stats} provides summary statistics of the molecules that make up the dataset. The drug-like molecules from AICures are generally medium-sized organic compounds, containing an average of 44.4 atoms (24.9 heavy atoms), up to a maximum of 181 atoms (91 heavy atoms). They contain a large variance in flexibility, as demonstrated by the mean (6.5) and maximum (53) number of rotatable bonds. 15\% (45,712) of the molecules have specified stereochemistry, while 27\% (83,326) have stereocenters but may or may not have specified stereochemistry. The QM9 dataset is limited to 9 heavy atoms (29 total atoms), with a much smaller molecular mass and few rotatable bonds. 72\% (95,734) of the species have specified stereochemistry. 

Table \ref{tab:target_stats} summarizes the experimental properties in the GEOM dataset from the AICures dataset. Of note is data for the inhibition of SARS-CoV-2, and for the specific inhibition of the SARS-CoV-2 3CL protease. The 3CL protease shares 96\% sequence similarity with its SARS-CoV 3CL counterpart \cite{cov_3cl_96_pct}, for which there is significantly more experimental data. The similarity of the two proteases means that CoV-2 models may benefit from pre-training with CoV data, so GEOM can also be used to benchmark transfer learning methods. Another target of interest is the SARS-CoV PL protease \cite{aid_485353, aid_652038}. The dataset also contains molecules screened for growth inhibition of \textit{E. Coli} and \textit{Pseudomonas aeruginosa}, both of which can cause secondary infections in COVID-19 patients.

Table \ref{tab:molecule_net} shows the species from MoleculeNet \cite{Wu2018MoleculeNet} that are included in GEOM. We used every compound from the physical chemistry and physiology categories. These molecules have experimental data for three physical chemistry tasks and 659 physiology tasks. The latter include blood-brain barrier penetration, qualitative toxicity, and whether a drug fails in clinical trials due to toxicity. GEOM also contains the BACE dataset \cite{subramanian2016computational}, which is part of the biophysics category of MoleculeNET. Each BACE molecule has an experimental binding affinity for human $\beta$-secretase 1 (BACE-1). The remaining biophysics datasets were excluded because of size, and because the AICures drug dataset is already sufficiently large. The ``recovered'' column in Table \ref{tab:molecule_net} shows that vacuum conformer-rotamer ensembles (CREs) were generated for over 98\% of the molecules in each dataset other than SIDER. CREST CREs were also generated with an implicit solvent model of water for 99.9\% of the BACE compounds. As mentioned above, these conformers were further annotated with single-point DFT energies and xTB quasi-harmonic free energies.

GEOM contains vacuum CREs for 98\% of the original molecules in all but one of the datasets within MoleculeNet. This means that future models using the CREs can be benchmarked against past predictions from 2D and single-conformer models \cite{Wu2018MoleculeNet}. Care should still be taken when making such comparisons, as the missing molecules have similar characteristics, and may therefore bias the resulting data. For example, many missing compounds are extremely flexible. For most of these compounds, the CREST calculations ran for several days with 40 cores and did not finish. Other missing compounds failed during initial xTB optimization, often because of unusual topologies; this was most common in the SIDER dataset.

\section*{Methods}

\subsection*{CREST}
Generation of conformers ranked by energy is computationally complex. Many exhaustive, stochastic, and Bayesian methods have been developed to generate conformers \cite{balloon_1, balloon_2, confab, frog2, moe, omega, rdkit, chan2019bayesian}. The exhaustive method is to enumerate all the possible rotations around every bond, but this approach has prohibitive exponential scaling with the number of rotatable bonds \cite{Schwab2010,OBoyle2011}. Stochastic algorithms available in cheminformatics packages such as RDKit \cite{rdkit} suffer from two flaws. First, they explore conformational space very sparsely through a combination of pre-defined distances and stochastic samples \cite{Spellmeyer1997} and can miss many low-energy conformations. Second, in most standalone applications, conformer energies are determined with classical force fields, which are rather inaccurate \cite{Kanal2018}. Enhanced molecular dynamics simulations, such as metadynamics (MTD), can sample conformational space more exhaustively, but need to evaluate an energy function many times. \textit{Ab initio} methods, such as DFT, can assign energies to conformers more accurately than force fields, but are also orders of magnitude more computationally demanding.

An efficient balance between speed and accuracy is offered by the newly developed CREST software \cite{CREST}. This program uses semi-empirical tight-binding DFT to calculate the energy. The predicted energies are significantly more accurate than classical force fields, accounting for electronic effects, rare functional groups, and bond-breaking/formation of labile bonds, but are computationally less demanding than full DFT. Moreover, the search algorithm is based on MTD, a well-established thermodynamic sampling approach that can efficiently explore the low-energy search space. Finally, the CREST software identifies and groups rotamers, conformers that are identical except for atom re-indexing. It then assigns each conformer a probability through 
\begin{align}
p_i^{\mathrm{CREST}} = \frac{d_i \ \mathrm{exp}(-E_i / k_\mathrm{B} T) }{\sum_j  d_j \  \mathrm{exp}(-E_j / k_\mathrm{B} T)}. \label{eq:p_crest}
\end{align}
Here $p_i$ is the statistical weight of the $i^{\mathrm{th}}$ conformer, $d_i$ is its degeneracy (i.e., how many chemically and permutationally equivalent rotamers correspond to the same conformer), $E_i$ is its energy, $k_{\mathrm{B}} $ is the Boltzmann constant, $T$ is the temperature, and the sum is over all conformers. Equation (\ref{eq:p_crest}) is an approximation to the true probability, $p_i \propto \mathrm{exp}(-G_i/ k_\mathrm{B} T)$, where $G$ is the free energy [Eqs. (\ref{eq:p_censo})-(\ref{eq:gibbs})]. The solvation free energy can be incorporated into $E$ with a solvent model, but the translation, rotation, and vibrational free energies are missing. The addition of these terms is discussed below. 

To generate conformers and rotamers, CREST takes a geometry as input and uses its flexibility to determine an MTD simulation time $t_{\mathrm{max}}$ (between 5 and 200 ps). The initial structure is deformed by propagating Newton's equations of motion with an NVT thermostat \cite{gc_2} from time $t=0$ to $t_{\mathrm{max}}$. The potential at each time step is given by the sum of the GFN2-xTB potential energy and a bias potential,
\begin{align}
    V_{\mathrm{bias}} = \sum_{i}^{n} k_i \mathrm{exp}(-\alpha_i \Delta_i^2), \label{eq:v_bias}
\end{align}
which forces the molecule into new conformations. The collective variables $\Delta_i$ are the root-mean-square displacements (RMSDs) of the structure with respect to the $i^{\mathrm{th}}$ reference structure, $n$ is the number of reference structures, $k_i$ is the pushing strength and $\alpha_i$ determines the potentials' shapes. A new reference structure from the trajectory is added to $V_{\mathrm{bias}}$ every 1.0 ps, driving the molecule to explore new conformations.  Different molecules require different $(k_i, \alpha_i)$ pairs to produce best results, so twelve different MTD runs are used with different settings for the $V_{\mathrm{bias}}$ parameters.

Conformers are defined by rotation about dihedral angles. In MTD simulations with RMSD collective variables, the biasing potential in Eq. (\ref{eq:v_bias}) generates energy for overcoming torsional barriers. Since it takes less energy to cross a rotational barrier than to break a covalent bond, the biasing term leads to exploration of conformational space through rotation, rather than to trivial fragmentation \cite{CREST, mtd}. Indeed, the bare energy without the biasing term keeps the molecule from exploring ultra-high energy regions, and thus reduces the size of the $3N-6$-dimensional PES to be explored, where $N$ is the number of atoms. This also makes it more efficient at finding accessible minima than an exhaustive enumeration of dihedral angles, since the latter would include high-energy, thermally inaccessible structures.

Geometries from the MTD runs are then optimized with GFN2-xTB. Conformers are identified as structures with $\Delta E > E_{\mathrm{thr}}$, $\text{RMSD} > \text{RMSD}_{\mathrm{thr}}$, and $\Delta B_e > B_{\mathrm{thr}}$, where $\Delta E$ is the energy difference between structures, $\Delta B_e$ is the difference in their rotational constants, and thr denotes a threshold value. Rotamers are identified through $\Delta E < E_{\mathrm{thr}}$, $\text{RMSD} > \text{RMSD}_{\mathrm{thr}}$, and $\Delta B_e < B_{\mathrm{thr}}$. Duplicates are identified through $\Delta E < E_{\mathrm{thr}}$, $\text{RMSD} < \text{RMSD}_{\mathrm{thr}}$, and $\Delta B_e < B_{\mathrm{thr}}$. The defaults, which are used in this work, are $E_{\mathrm{thr}}=0.1$ kcal/mol, $\mathrm{RMSD}_{\mathrm{thr}}=0.125$ \AA, and $B_{\mathrm{thr}}=15.0$ MHz. Conformers and rotamers are added to the CRE and duplicates are discarded.

If a new conformer has a lower energy than the input structure, the procedure is restarted using the conformer as input, and the resulting structures are added to the CRE. The procedure is restarted between one and five times. The three conformers of lowest energy then undergo two normal molecular dynamics (MD) simulations at 400 K and 500 K. These are used to sample low-energy barrier crossings, such as simple torsional motions, which are need to identify the remaining rotamers. Conformers and rotamers are once again identified and added to the CRE. All accumulated structures are then used as inputs to a genetic Z-matrix crossing algorithm \cite{gc_1, gc_2}, the results of which are also added to the CRE. All geometries accumulated throughout the sampling process are optimized with a tight convergence threshold, identified as conformers, rotamers or duplicates, and sorted to yield the final set of structures. The process is restarted after the regular MD runs or the tight optimization if any conformers have lower energy than the input, with no limit to the number of restarts. The final CRE contains conformers and rotamers up to a maximum energy $E_{\mathrm{win}}$. The default $E_{\mathrm{win}} = 6.0$ kcal/mol provides a safety net around errors in the xTB energies, as only conformers with $E \lessapprox 2.5$ kcal/mol have significant population at room temperature.


CREST generates ensembles with good coverage of the true CREs. For example, Ref. \cite{CREST} compared the experimental conformations of gas-phase citronellal, inferred through microwave spectroscopy \cite{domingos2016flexibility}, with computational predictions. Each of the 15 lowest-energy experimental conformers was found in the CREST ensemble. The $^{1}\mathrm{H}$-NMR spectrum was then computed in chloroform using CREST conformers, together with DFT for energy re-ranking and computation of the coupling and shielding constants. The spectrum with the ensemble matched experiment far better than with only one conformer \cite{CREST}. Investigation of macrocycles, a protonated peptide, metal-organic systems, and the 1-Naphthol dimer yielded similarly good results. 

\subsection*{DFT}
CREST offers an excellent balance between cost and accuracy for generating an initial CRE. The GFN2-xTB method is fast enough to be used in long MTD runs, and its conformational energies are accurate to within 2 kcal/mol (see Technical Validation). The number of energy and force calculations can easily reach into the millions for a single CREST run, making full DFT prohibitive and xTB quite practical. Further, the CREST safety window of 6.0 kcal/mol ensures that the vast majority of accessible conformers should be present in the CRE. However, the typical xTB errors of 2 kcal/mol are too large for the accurate ranking of the conformers by statistical weight. This is because $p$ is exponential in $\Delta E/k_{\mathrm{B}}T$, and at room temperature $k_{\mathrm{B}}T = 0.59$ kcal/mol, which is 3.4 times smaller than the average error. Further, the weights do not take into account the zero-point energy or the roto-translational and vibrational entropy (see below). Each of these contributions to the free energy is conformation-dependent, and can lead to non-negligible changes in statistical weight.

DFT can be used to optimize conformers and compute their relaxed energies. However, each ensemble can contain hundreds of conformers, which makes DFT optimization extremely resource-intensive. Further, a Hessian calculation is required to compute the zero-point energy and entropic corrections to the free energy. Such calculations are among the most computationally demanding in quantum chemistry. Thus a full DFT optimization of each ensemble, together with an accurate free energy calculation, is a daunting task.

To address these issues, the developers of CREST recently introduced the CENSO program \cite{grimme2021efficient}. CENSO uses a series of optimizations at increasingly accurate levels of DFT theory. The free energy cutoff for discarding conformers is reduced at each stage, leading to fewer conformers in each successive round. Further, CENSO uses the recently developed r2scan-3c meta-GGA functional \cite{grimme2021r2scan} for the final optimization. r2scan-3c with the custom-made mTZVPP basis set is extremely accurate, yielding conformational energies that are within 0.3 kcal/mol of the CCSD(T) complete basis set limit \cite{grimme2021r2scan}. It is also quite affordable given its accuracy, with a cost that is 100-1000 times lower than hybrid functionals with large basis sets \cite{grimme2021r2scan}. The optimization is further accelerated by discarding duplicate conformers and high-energy geometries that are close to converged \cite{grimme2021efficient}. Lastly, CENSO computes entropic and zero-point corrections using the new biased Hessian method \cite{spicher2021single}. This technique uses xTB, which is quite computationally affordable, together with an extra biasing potential. The biasing potential accounts for energy differences between xTB and DFT, which allows xTB Hessians to be computed for DFT-optimized geometries. 

The statistical weight computed by CENSO for the $i^{\mathrm{th}}$ conformer is 
\begin{align}
p_i^{\mathrm{CENSO}} = \frac{\mathrm{exp}(-G_i / k_\mathrm{B} T)}{\sum_j  \mathrm{exp}(-G_j / k_\mathrm{B} T)}, \label{eq:p_censo}
\end{align}
where $G_i$ is the conformation-dependent free energy. Note that unlike CREST, CENSO does not include rotamer degeneracy in the calculation of $p$. The reason is that accounting for all rotamers, though attempted by CREST, is still a difficult task, and the difference in rotamers among different conformers should be small. The free energy is given by \cite{grimme2021efficient}:
\begin{align}
    G_i = E^{(i)}_{\mathrm{gas}} + \delta G^{(i)}_{\mathrm{solv}}(T) + G^{(i)}_{\mathrm{trv}}(T). \label{eq:gibbs}
\end{align}
Here $E_{\mathrm{gas}}$ is the gas phase energy, $\delta G_{\mathrm{solv}}(T)$ is the solvation free energy, and $G_{\mathrm{trv}}(T)$ is the free energy due to translation, rotation and vibration. $\delta G_{\mathrm{solv}}$ can be calculated with implicit solvent methods such as COSMO-RS \cite{klamt1995conductor, klamt1998refinement} or C-PCM \cite{barone1998quantum}. The solvation free energy predicted by r2scan-3c/COSMO-RS is typically accurate to within 0.5 kcal/mol \cite{grimme2021efficient}. Given the Hessian matrix and the associated normal modes, $G_{\mathrm{trv}}(T)$ can be computed within the standard modified rigid-rotor harmonic-oscillator approximation \cite{grimme2012supramolecular}. This term can be predicted quite accurately, with sub-chemical accuracy attainable even for semi-empirical methods \cite{grimme2021efficient}. 

CENSO qualitatively reproduces the optical rotation of organic molecules measured in solution, which is a challenging task that depends sensitively on the CRE \cite{grimme2021efficient}. Further, it makes very accurate predictions of the octanol-water partition coefficients and $\mathrm{p}K_{\mathrm{a}}$ values of various organic molecules \cite{grimme2021efficient}. Conformers and statistical weights generated by CENSO are thus quite reliable.

In this work we apply CENSO to 534 species, yielding the highest-accuracy ensembles ever generated for drug-like molecules. Calculations are performed in implicit water solvent for 35\% of the molecules in the BACE dataset \cite{subramanian2016computational, Wu2018MoleculeNet}, which contains experimental binding affinities for inhibitors of BACE-1 (Table \ref{tab:molecule_net}). Binding affinity models that incorporate CREs can be trained with this data. Models trained on a single conformer can also benefit from the CENSO ensembles. Since many of the drug-like molecules are quite flexible, the typical approach of optimizing a single force field conformer with DFT is likely to miss the true lowest-energy structure. Thus the lowest-energy CENSO structures are far more reliable inputs to single-conformer models. Lastly, the ensembles can be used for transfer learning (TL), so that generative models trained on the large CREST dataset can be fine-tuned with the CENSO data. 

In addition to the fully optimized CREs, we provide single-point DFT energies for all 1.3 million CREST conformers in 1,511 out of 1,513 BACE species (99.9\%). We also provide xTB vibrational frequencies to complete the calculation of $G$. Together, these calculations give statistical weights that are much more accurate than those of CREST, and somewhat less accurate than CENSO. Since nearly all BACE species have single-point calculations, future binding affinity models using the re-ranked CREs can be benchmarked against predictions from past 2D and 3D models \cite{Wu2018MoleculeNet}. All geometries with DFT energies are also annotated with DFT dipole moments, partial charges, and molecular orbital energies. This data can be used for multi-task learning to improve TL for conformer generation.

\subsection*{Conformer generation}
\subsubsection*{SMILES pre-processing}
SMILES strings from the QM9 dataset were used as given. SMILES strings and properties of the drug-like molecules were accessed from Refs. \cite{aicures_dsets, pseudomonas} (original sources are \cite{ellinger, touret2020vitro, diamond_light, aid_binarized_sars_source, aid_485353, aid_652038, ecoli_1, ecoli_2, pseudomonas}). Each SMILES string was converted to its canonical form using RDKit. This allowed us to assign multiple properties from multiple sources to a single species, even if different non-canonical SMILES strings were used in the original sources.

3.9\% of the drug molecules accessed (11,886 total) were given as clusters, either with a counterbalancing ion (e.g. ``.[Na+]'', ``.[Cl-]'') or with an acid to represent the protonated salt (e.g. ``.Cl''). For non acid-base clusters we identified the compound of interest as the heaviest component of the cluster. For the acid/base SMILES strings, used reaction SMARTS in RDKit to generate the protonated molecule and counterion. This product SMILES was used in place of the original SMILES. Original SMILES strings are available in the dataset with the key \texttt{uncleaned\_smiles} (see Ref. \cite{geom_git} for details). Not only does de-salting identify the drug-like compound in each cluster and correct its ionization state, it also homogenizes the molecular representations in the drug datasets. For MoleculeNet we also selected the heaviest component from each cluster SMILES, but did not perform protonation. 

\subsubsection*{Initial structure generation}
To generate conformers with CREST one must provide an initial guess geometry, ideally optimized at the same level of theory as the simulation (GFN2-xTB). For the drug molecules we therefore used RDKit to generate initial conformers from SMILES strings, optimized each conformer with GFN2-xTB, and used the lowest energy conformer as input to CREST. 

Conformers were generated in RDKit using the \texttt{EmbedMultipleConfs} command with 50 conformers (\texttt{numConfs = 50}), a pruning threshold of similar conformers of 0.01 \AA \ (\texttt{pruneRmsThresh = 0.01}), a maximum of five embedding attempts per conformer (\texttt{maxAttempts = 5}), coordinate initialization from the eigenvalues of the distance matrix (\texttt{useRandomCoords = False}), and a random seed. If no conformers were successfully generated then \texttt{numConfs} was increased to 500. Each conformer was then optimized with the MMFF force field \cite{mmff} in RDKit using the default arguments. Duplicate conformers, identified as those with an RMSD below 0.1 \AA, were removed after optimization. Optimization was skipped for any molecules with \textit{cis}/\textit{trans} stereochemistry (indicated by ``\textbackslash'' or ``/'' in the SMILES string), as such stereochemistry is not always maintained during RDKit optimization.

The ten MMFF-optimized conformers with the lowest energy were further optimized with xTB using Orca 4.2.0 \cite{orca_1, orca_2}. The conformer with the lowest xTB energy was selected as the seed geometry for CREST. The QM9 molecules are already optimized with DFT, and so in principle did not need to be optimized further for CREST. However, since it is recommended to seed CREST with a structure optimized at the GFN2-xTB level of theory, we re-optimized each QM9 geometry with xTB before using it in CREST.

\subsubsection*{CREST simulation}

A single xTB-optimized structure was used as input to the CREST simulation of each species. Default values were used for all CREST arguments, except for the charge of each geometry. CREST runs on the AICures drug dataset took an average of 2.8 hours of wall time on 32 cores on Knights Landing (KNL) nodes (89.1 core hours), and 0.63 hours on 13 cores on Cascade Lake and Sky Lake nodes (8.2 core hours). QM9 jobs were only performed on the latter two nodes, and took an average of 0.04 wall hours on 13 cores (0.5 core hours). 13 million KNL core hours and 1.2 million Cascade Lake/Sky Lake core hours were used in total.

CREST calculations on MoleculeNet species were run across several compute clusters, each with various node types and different core counts per node. KNL nodes were not used. Excluding species already present in the AICures dataset, each MoleculeNet job took 6.3 hours of wall time using 18.1 cores on average. These values are skewed by extremely flexible molecules whose CREST jobs took several days to finish: the \textit{median} wall time was 1.4 hours, and the median core count was 12.0. 1.5 million CPU hours were used in total.

\subsubsection*{Graph re-identification}
It was necessary to re-identify the graph of each conformer generated by CREST, for the following reasons. First, stereochemistry may not have been specified in the original SMILES string, but necessarily existed in each of the generated 3D structures. Second, reactivity such as dissociation or tautomerization may have occurred in the CREST simulations (CREST has specific commands to generate tautomers, but they were not used here). This would also lead to conformers with different graphs.

To re-identify the graphs we used \texttt{xyz2mol} \cite{xyz2mol} (code accessed from \cite{xyz2mol_git}) to generate an RDKit \texttt{mol} object. These \texttt{mol} objects were used to assign graph features to each conformer (see Data Records) . It should be noted that \texttt{xyz2mol} sometimes assigned resonance structure graphs instead of the original graphs. 
In some cases this caused different conformers of the same species to have different graphs. This happened, for example, when the conformers had different \textit{cis/trans} isomerism about a double bond that was only present because of the resonance structure used (see the RDKit tutorial in Ref. \cite{geom_git}). This is conceptually different from species whose conformer graphs differ because of reactivity. One may want to distinguish these two cases when analyzing the conformer \texttt{mol} objects.

\subsubsection*{CENSO simulation}
534 molecules from the BACE dataset (35\%) were optimized with CENSO. Initial CREs were generated with CREST using the ALPB model for water \cite{ehlert2021robust}. The CREs were refined with CENSO 1.1.2, using Orca 5.0.1 \cite{neese2020orca} to perform the DFT calculations. The C-PCM \cite{barone1998quantum} model of water was used for DFT and the ALPB model was used for xTB. Conformer and rotamer duplicates were removed throughout the optimization using CREST (\texttt{crestcheck="on"}). Default values were used for all other parameters. We used the same clusters and nodes for CENSO as for CREST with MoleculeNet species. The average CENSO job took 1 day and 4 hours of wall time using 54 cores. 781,000 CPU hours were used in total.

\subsubsection*{Single point calculations}
We performed single-point DFT calculations on all CREST conformers in the BACE dataset without further optimization. We used Orca version 5.0.2 and the same level of theory as in CENSO optimization (r2scan-3c functional, mTZVPP basis, C-PCM model of water, and default grid 2). The average run took 6.4 minutes of wall time using 8 cores. Calculations took a total of 1.1 million CPU hours for 1.3 million conformers.

\subsubsection*{Hessian calculations}
We performed Hessian calculations on all CREST conformers in the BACE dataset, using xTB with the ALPB model for water. The average run took 41 seconds of wall time using 4 cores. Calculations took a total of 63,000 CPU hours for 1.3 million conformers.

\subsection*{Conformational property prediction}

The GEOM dataset is significant because it allows for the training of conformer-based property predictors and generative models to predict new conformations. The first application will be explored in a future publication \cite{covid_paper}. The second application is necessary for using conformer-based ML models in practice, since generating CREST structures from scratch is too costly for the virtual screening of new species. Such work is already underway \cite{geom_benchmark}, paving the way for \texttt{graph} $\rightarrow$ \texttt{conformer ensemble} $\rightarrow$ \texttt{property} models that can be trained end-to-end. Here we give an example of a simpler application in the same vein, benchmarking methods to predict \textit{summary statistics} of each conformer ensemble, rather than the conformers themselves. Our proposed tasks are similar to the benchmark QM9 tasks, which measure a model's ability to predict properties that are uniquely determined by geometry. Here, since we provide conformer ensembles for each species, we measure a model's ability to predict properties defined by the ensemble. Because one chemical graph spawns a unique conformer ensemble, these tasks are also a metric of the performance of graph-based models to infer properties mediated through conformational flexibility. 

\begin{figure}[H]
    \centering
    \begin{tikzpicture}
    \node[inner sep=0pt] at (-2.45,0)
    {\includegraphics[width=0.23\textwidth,trim={0 0 15cm 0},clip]{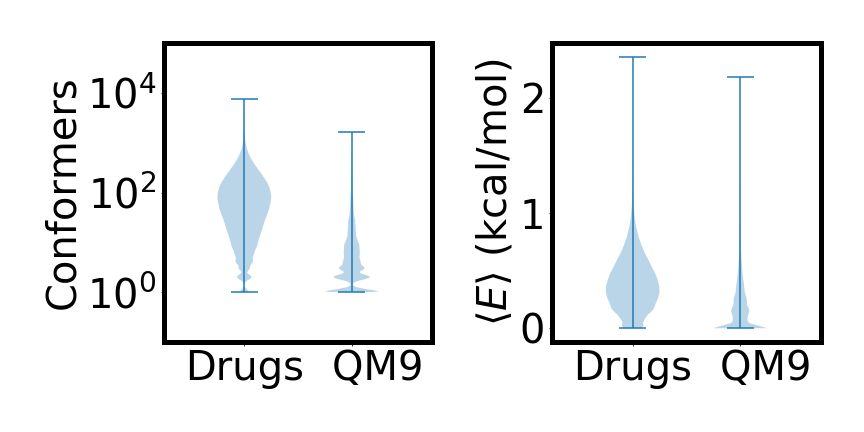}};
    \node[inner sep=0pt] at (1.8,0)
    {\includegraphics[width=0.22\textwidth,trim={16cm 0 0 0},clip]{violin_side_by_side_0.png}};
    \node[inner sep=0pt] at (-0.3,-4)
        {\includegraphics[width=0.45\textwidth]{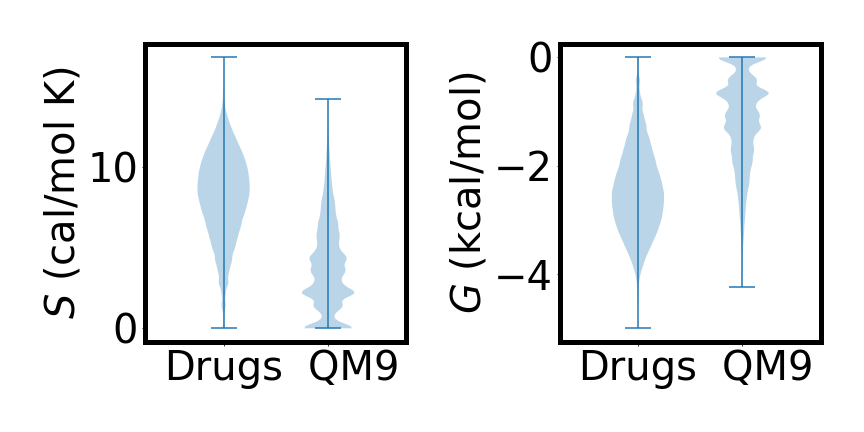}};
    \end{tikzpicture}
    \caption{Violin plots of CREST-based statistics for the QM9 and AICures drug datasets.}
    \label{fig:data_stats}

\end{figure}

We trained different models to predict three quantities related to conformational information. A summary of these quantities can be found in Table \ref{tab:data_stats} and Fig. \ref{fig:data_stats}. The first quantity is the conformational free energy, $G=- TS$, where the ensemble entropy is $S= -R \sum_i p_i \ \mathrm{log} \ p_i$ \cite{CREST}. Here the sum is over the statistical probabilities $p_i$ of the $i^{\mathrm{th}}$ conformer, and $R$ is the gas constant. The conformational entropy is a measure of the conformational degrees of freedom available to a molecule. A molecule with only one conformer has an entropy of exactly 0, while a molecule with equal statistical weight for an infinite number of conformers has infinite conformational entropy. The conformational Gibbs free energy is an important quantity for predicting the binding affinity of a drug to a target. The affinity is determined by the change in Gibbs free energy of the molecule and protein upon binding, which includes the loss of molecular conformational free energy \cite{frederick2007conformational}. The second quantity is the average conformational energy. The average energy is given by $\langle E \rangle = \sum_i p_i E_i$, where $E_i$ is the energy of the $i^{\mathrm{th}}$ conformer. Each energy is defined with respect to the lowest-energy conformer. The third quantity is the number of unique conformers for a given molecule, as predicted by CREST within the default maximum energy window \cite{CREST}. 

We trained a kernel ridge regression (KRR) model \cite{murphy2012machine}, a random forest \cite{breiman2001random}, and three different neural networks to predict conformer properties. The random forest, KRR and feed-forward neural network (FFNN) models were trained on Morgan fingerprints \cite{morgan} generated through RDKit. Two different message-passing neural networks \cite{mpnn} were trained. The first, called ChemProp, has achieved state-of-the-art performance on a number of benchmarks \cite{chemprop}. The second is based on the SchNet force field model \cite{schnet_1, schnet_2}. We call it SchNetFeatures, as it learns from 3D geometries using the SchNet architecture, but also incorporates graph-based node and bond features. The SchNetFeatures models were trained on the highest-probability conformer of each species.

100,000 species were sampled randomly from the AICures drug subset of GEOM. We used the same 60-20-20 train-validation-test split for each model. The splits, trained models, and log files can be found with the GEOM dataset at \cite{dataverse_models}, under the heading ``synthetic''. Hyperparameters were optimized for each model type and for each task using the hyperopt package \cite{hyperopt}. Details of the hyperparameter searches, optimal parameters, and network architectures can be found in \cite{dataverse_models}. Source code is available at \cite{nff}.

Results are shown in Table \ref{tab:results}. ChemProp and SchNetFeatures are the strongest models overall, followed in order by FFNN, KRR, and random forest. Of the three models that use fixed 2D fingerprints, we see that the FFNN is best able to map these non-learnable representations to properties. ChemProp has the added flexibility of learning an ideal molecular representation directly from the graph, and so performs even better than the FFNN. The SchNetFeatures model retains this flexibility while incorporating extra information from one 3D structure. Compared to ChemProp, its prediction error is 10\% lower for $G$, nearly equal for $\braket{E}$, and 5\% lower for ln(unique conformers).  This is not surprising, as the ensemble properties are mainly determined by molecular flexibility, which is a function of the graph through the number of rotatable bonds. A single 3D geometry would not provide extra information about this flexibility.

We see that various models can accurately predict conformer properties when trained on the GEOM dataset. With access to the dataset, researchers will therefore be able to predict results of expensive simulations without performing them directly. This has implications beyond ensemble-averaged properties, as generative models trained on the GEOM dataset will also be able to produce the conformers themselves \cite{geom_benchmark}.

\section*{Data Records}

The dataset is available online at \cite{geom_dataverse}, and detailed tutorials for loading and analyzing the data can be found at \cite{geom_git}.

The data is available either through MessagePack \cite{msg_pack}, a language-agnostic binary serialization format, or through Python pickle files. There are two MessagePack files for the AICures drug dataset and two for QM9. Each of the two files contains a dictionary, where the keys are SMILES strings and the values are sub-dictionaries. In the file with suffix \texttt{crude}, the sub-dictionaries contain both species-level information (experimental binding data, average conformer energy, etc.) and a list of dictionaries for each conformer. Each conformer dictionary has its own conformer-level information (geometry, energy, degeneracy, etc.). In the file with suffix \texttt{featurized}, each conformer dictionary contains information about its molecular graph.

The Python pickle files are organized in a different fashion. The main folder is divided into sub-folders for QM9, AICures, and MoleculeNet data, plus separate folders for BACE calculations in water and with CENSO. Each sub-folder contains one pickle file for each species. Each pickle file contains both summary information and conformer information for its species. Each conformer is stored as an RDKit \texttt{mol} object, so that it contains both the geometry and graph features. One may only want to load the pickle files of species with specific properties (e.g., those with experimental data for SARS-CoV-2 inhibition); for this one can use the summary \texttt{JSON} file. This file contains all summary information along with the path to the pickle file, but without the list of conformers. It is therefore lightweight and quick to load, and can be used to choose species before loading their pickles.

\section*{Technical Validation}
The quality of the data was validated in three different ways. First, we checked that the conformer data was accurately parsed from the CREST calculations. To do so we randomly sampled one conformer from 20 different species and manually confirmed that its data matched the data in the CREST output files. 

Second, we re-identified the graphs of the conformers generated by CREST using \texttt{xyz2mol}. The graph re-attribution procedure succeeded for 88.4\% of the QM9 molecules and 94.7\% of the drug molecules, recovering the original molecular graph that was used to generate each conformer. Note that to compare graphs we removed stereochemical indicators from the original and the re-generated graph. This was done because of cases in which stereochemistry was not specified originally but was specified in the generated conformers. All of the failed QM9 graphs underwent some sort of reaction, which can be explained by the presence of highly strained and unstable molecules. However, manual inspection of 53 cases in the AICures drug dataset suggests that 70\% of the drug graphs failed only because of poor handling of resonance forms by \texttt{xyz2mol} (see above). This means that the original graph was likely recovered for 98.4\% of all drugs. 21\% of cases failed because of tautomerization (1\% of all cases), and 9.4\% failed because of a different reaction (usually dissociation or ring formation; 0.5\% of all cases). The high success rate of the graph re-identification indicates that, in the vast majority of cases, the geometries generated by CREST were actual conformers of the species.

\begin{figure}[H]
    \centering
    \includegraphics[width=\textwidth,trim={0 0 0 0},clip]{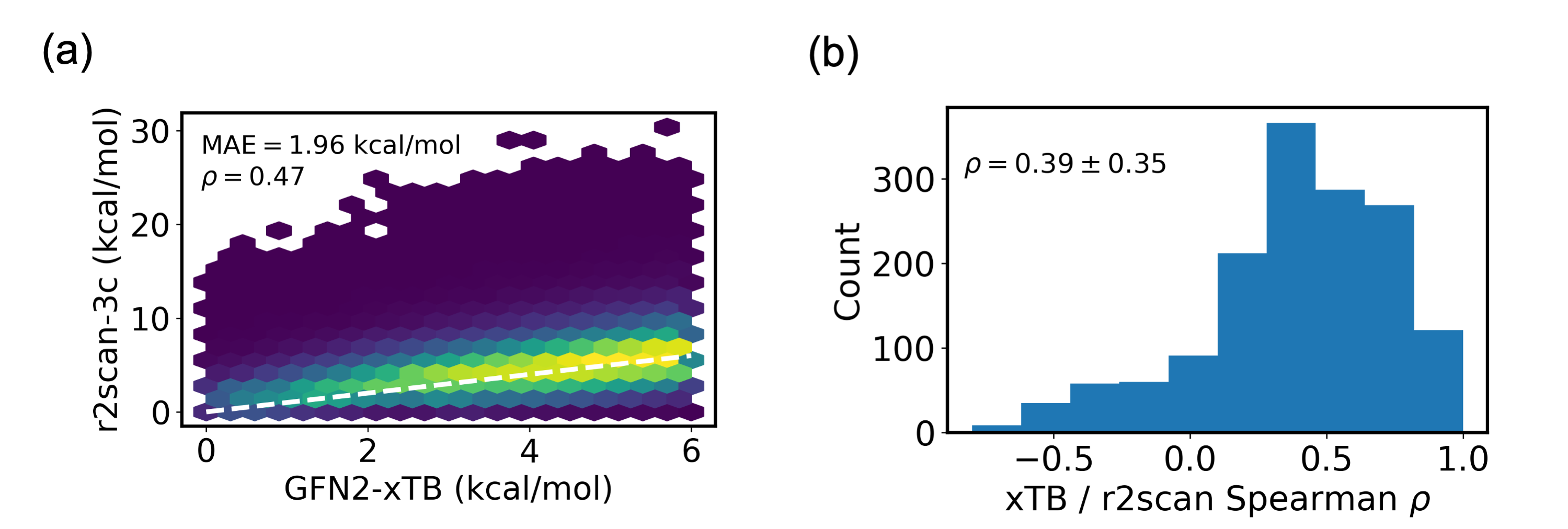}

    \caption{Comparison of GFN2-xTB (CREST) energies and single-point r2scan-3c (DFT) energies. (a) xTB vs. r2scan-3c energies for all geometries in the BACE-1 dataset. The ideal correlation is shown with a dashed white line. (b) Distribution of Spearman rank correlation coefficients $\rho$, measuring the accuracy of xTB energy ranking for each of the ensembles. }
    \label{fig:xtb_vs_sp}
    
\end{figure}

Third, we compared the CREST energies and coordinates to those from higher levels of theory. 
Figure \ref{fig:xtb_vs_sp} compares the GFN2-xTB calculations of CREST with single-point r2scan-3c calculations, both performed in water for 1,511 species in the BACE dataset. Panel (a) shows the relative energies of the two methods. The mean absolute error (MAE) of xTB is 1.96 kcal/mol, which is similar to reported values in conformational energy benchmarks \cite{bannwarth2019gfn2}. The ranking accuracy can be measured with the Spearman correlation coefficient $\rho$, which lies between 1 and $-1$ (perfect correlation and anti-correlation, respectively). The Spearman coefficient is 0.47 when using all geometries from all species. However, it is more meaningful to judge the energy rankings among different conformers in a single species. Computing $\rho$ separately for each species yields the distribution in panel (b). The distribution of $\rho$ is quite wide, with an average value of 0.39 and a standard deviation of 0.35. The mean value of $\rho$ indicates moderate correlation between the methods. The correlation is significantly better than for classical force fields such as MMFF94 \cite{mmff}, UFF \cite{rappe1992uff}, and GAFF \cite{wang2004development}: For instance, the median $\rho$ between MMFF94 and single-point DFT for drug-like molecules is between $-0.1$ and $-0.45$, meaning that the two methods are actually weakly \textit{anti-correlated} (Supporting Information of Ref. \cite{Kanal2018}).

\begin{figure}[H]
    \centering
    \includegraphics[width=\textwidth,trim={0 0 0 0},clip]{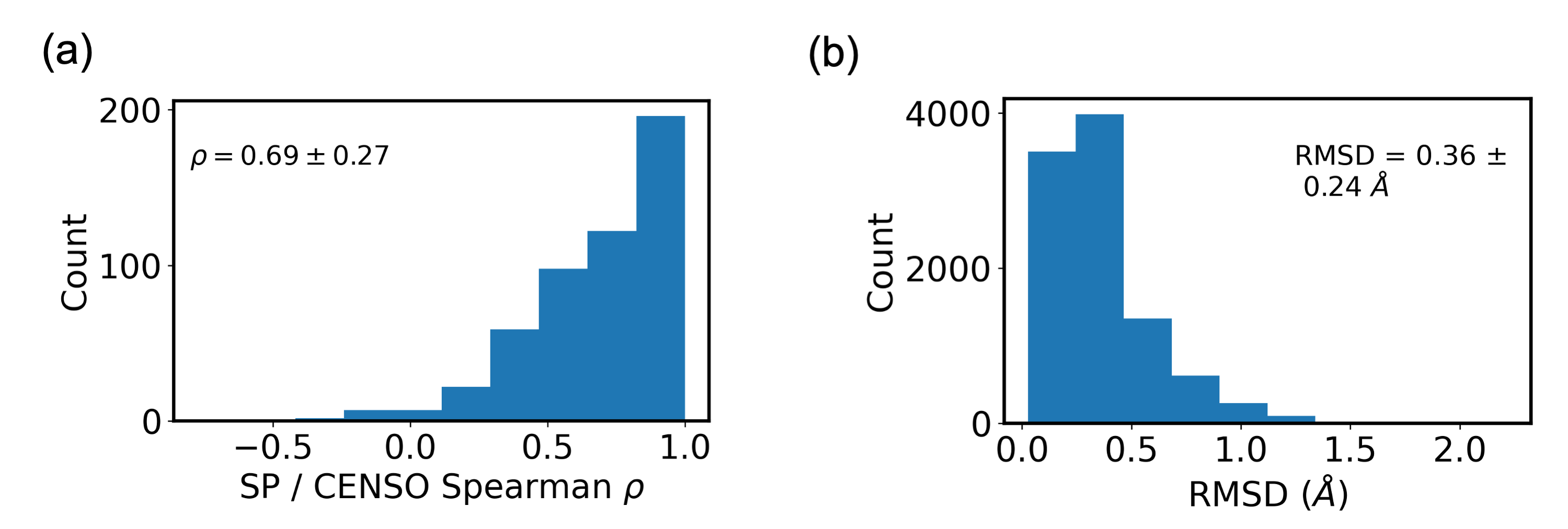}

    \caption{Comparison of CENSO and single-point DFT calculations. (a) Distribution of Spearman coefficients, measuring the accuracy of single-point ranking for each of the ensembles. (b) RMSDs between CREST geometries and DFT-optimized geometries. }
    
    \label{fig:sp_vs_censo}

\end{figure}

Figure \ref{fig:sp_vs_censo} compares single-point DFT calculations on CREST geometries (``SP'') with DFT results on fully optimized geometries (``CENSO''). Panel (a) shows the distribution of $\rho$ for conformer energies. The average Spearman correlation is 0.69 and the standard deviation is 0.27, indicating good agreement between the two methods. Indeed, the MAE between optimized and single-point relative energies is 0.54 kcal/mol, which is 3.6 times lower than the xTB error (the MAE of the absolute energy, equal to the average energy released after optimization, is 5.74 kcal/mol). Panel (b) shows that the geometries change very little during optimization, with a mean RMSD of only 0.36 \AA. This shows that the CREST geometries are quite good, thus validating the quality of the GEOM ensembles. The median RMSD among heavy atoms is 0.25 \AA; this is 2.4 times lower than the value of 0.6 \AA \ between MMFF94 and PM7 geometries \cite{stewart2013optimization} for drug-like molecules \cite{Kanal2018}. 

Similar comparisons can be made between CENSO geometries and their most similar CREST counterparts (i.e., the CREST geometry with the lowest RMSD relative to a CENSO geometry). These may not be the same as the CREST geometries used to seed the optimization. We have found that using the most similar geometries does not significantly affect the results; for example, the Spearman coefficient only climbs to 0.72 $\pm$ 0.27, while the RMSD only drops to 0.33 $\pm$ 0.19. Note also that the comparison of the methods only includes conformers with non-negligible weight after optimization ($\Delta G \leq 2.5$ kcal/mol), since CENSO discards high-energy conformers during optimization. Hence high-energy conformers were not fully optimized and thus not included in the comparison. 

\begin{figure}[H]
    \centering
    \includegraphics[width=\textwidth,trim={0 0 0 0},clip]{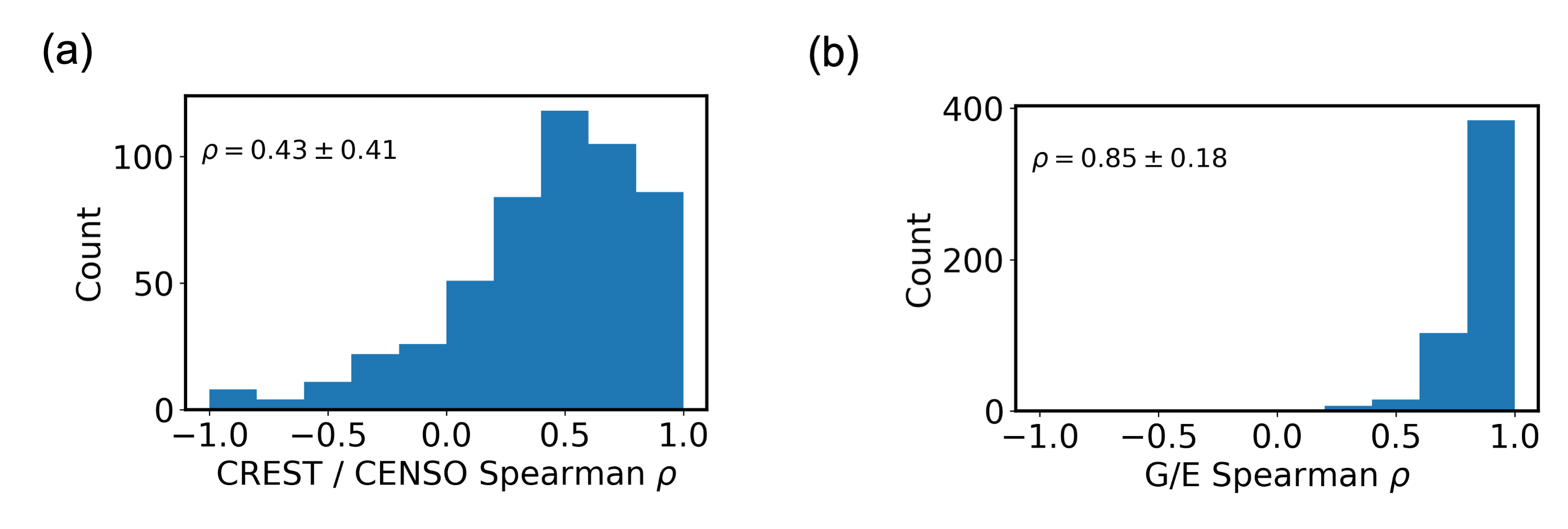}

    \caption{(a) Comparison of CENSO and CREST calculations. The distribution of Spearman coefficients shows the accuracy of CREST ranking for each of the ensembles. (b) Comparison of energy and free-energy ranking with CENSO. The distribution of Spearman coefficients shows the accuracy of energy ranking for each of the ensembles. }
    \label{fig:crest_vs_censo_and_free}

\end{figure}

Figure \ref{fig:crest_vs_censo_and_free} (a) compares the ordering of geometries with CREST and with CENSO. The Spearman correlation is $\rho=0.43 \pm 0.41$, which is similar to the correlation between CREST and single-point energies. This result should be interpreted with caution, however, since only the lowest-energy CENSO geometries are included in the comparison, whereas the rank correlation in Fig. \ref{fig:xtb_vs_sp} includes all CREST conformers. Lastly, Fig. \ref{fig:crest_vs_censo_and_free} (b) compares the ordering of CENSO geometries by energy and by free energy. The correlation is quite high ($\rho=0.85 \pm 0.18$), and the MAE between energies and free energies is only 0.33 kcal/mol. Hence energies alone can be quite good for ordering conformers by statistical weight. This also means that the statistical weight errors in GEOM are dominated by xTB errors, and that the quasi-harmonic errors are comparably negligible.

\section*{Usage Notes}

Researchers are encouraged to use the data-loading tutorials given in \cite{geom_git}. We suggest loading the data through the RDKit pickle files, as RDKit \texttt{mol} objects are easy to handle and their properties can be readily analyzed. The MessagePack files, while secure and accessible in all languages, represent graphs through their features rather than objects with built-in methods, and are thus more difficult to analyze. To train 3D-based models we suggest following the tutorial and \texttt{README} file in \cite{nff}.

\section*{Code availability}

Tutorials for loading the dataset  \cite{geom_git} and code for training 3D-based neural network models \cite{nff} are publicly available without restriction. CREST \cite{CREST_web} and xTB \cite{xtb_web} are both freely available online. CREST version 2.9 was used with xTB version 6.2.3 to generate the initial CREs. CENSO 1.1.2 was used with Orca 5.0.1 \cite{neese2020orca} and xTB 6.4.1 to refine the ensembles. Orca 5.0.2 was used for all single-point calculations. A race condition bug in version 5.0.1 meant that some CENSO energies were clearly incorrect (conformational energies above 1,000 kcal/mol), while some energy calculations failed to converge for reasonable geometries. Therefore, we discarded ensembles with failed energy calculations or conformational energy ranges exceeding 30 kcal/mol at any stage of the optimization. We also performed new single-point calculations on all converged CENSO geometries with Orca 5.0.2; 0.44\% of the energies were found to be incorrect and were replaced. 

\bibliography{main}

\section*{Acknowledgements}

The authors thank the XSEDE COVID-19 HPC Consortium, project CHE200039, for compute time. NASA Advanced Supercomputing (NAS) Division and LBNL National Energy Research Scientific Computing Center (NERSC), MIT Engaging cluster, Harvard Cannon cluster, and MIT Lincoln Lab Supercloud clusters are gratefully acknowledged for computational resources and support. We kindly thank Professor Eugene Shakhnovich (Harvard) for enlightening discussions. The authors also thank Christopher E. Henze (NASA) and Shane Canon and Laurie Stephey (NERSC) for technical discussions and computational support, MIT AI Cures (https://www.aicures.mit.edu/) for molecular datasets and Wujie Wang, Daniel Schwalbe Koda, Shi Jun Ang (MIT DMSE) for scientific discussions and access to computer code. Financial support from DARPA (Award HR00111920025) and MIT-IBM Watson AI Lab is acknowledged.

\section*{Author contributions statement}

R.G.B. conceived the project and S.A. performed the calculations. Both authors wrote and revised the manuscript.

\section*{Competing interests}
The authors declare no competing interests.

\section*{Figures \& Tables}

\setlength{\tabcolsep}{10pt}
\begin{table}[H]
    \centering
	\begin{tabular}{lcccc}
        \toprule
        &  AICures drug dataset (N=304,466) &  \\ \midrule
         & Mean & Standard deviation  & Maximum  \\ 
        \midrule
        Number of atoms & 44.4 & 11.3 & 181  \\
        Number of heavy atoms & 24.9 & 5.7  & 91 \\ 
        Molecular weight (amu) & 355.4  & 80.4  & 1549.7  \\
        Number of rotatable bonds & 6.5  & 3.0 & 53 \\
        Stereochemistry (specified) & 45,712 & - & -  \\
        Stereochemistry (all) & 83,326 & - &  -  \\
        \bottomrule
    \end{tabular}
    
	\begin{tabular}{lcccc}
        & QM9 dataset (N=133,258)  & \\ \midrule
        & Mean & Standard deviation  & Maximum \\ 
        \midrule
        Number of atoms & 18.0 & 3.0  & 29 \\
        Number of heavy atoms  & 8.8 & 0.51  & 9 \\ 
        Molecular weight (amu)  & 122.7 & 7.6 &  152.0 \\
        Number of rotatable bonds  & 2.2 & 1.6 & 8 \\
        Stereochemistry (specified)  & 95,734 & -  & - \\
        Stereochemistry (all) & 95,734 & - &  -  \\
        \bottomrule
    \end{tabular}

	\caption{Molecular descriptor statistics for the QM9 and AICures molecules in the GEOM dataset.}
	\label{tab:descriptor_stats}
\end{table}

\begin{table}[H]
	\vspace{6mm}
    \centering

    	\begin{tabular}{lcccc}
            \toprule
            Target & Species & Hits & Sources \\ 
            \midrule 
            SARS-CoV-2 & 5,832 & 101  & \cite{ellinger, touret2020vitro} \\
            SARS-CoV-2 3CL protease & 817 & 78  &  \cite{diamond_light}  \\
            SARS-CoV 3CL protease & 289,808 & 447 &  \cite{aid_binarized_sars_source}   \\
            SARS-CoV PL protease & 232,708 & 696 &  \cite{aid_485353, aid_652038}  \\ 
            \textit{E. Coli} & 2,186 & 111 & \cite{ecoli_1, ecoli_2}  \\
            \textit{Pseudomonas aeruginosa} & 1,968 & 48 & \cite{pseudomonas}   \\
            \bottomrule
        \end{tabular}
	\caption{Experimental data for GEOM species from AICures \cite{aicures}.}
	\label{tab:target_stats}
\end{table}

\begin{table}[H]
	\vspace{6mm}
    \centering

    	\begin{tabular}{l|cccccc}
            \toprule
            Category & Dataset & Property & Tasks & Species & Recovered & Sources \\ 
            \midrule 
            & ESOL & Water solubility & 1 & 1,113 & 99.6\% & \cite{delaney2004esol} \\
            Physical chemistry & FreeSolv & Hydration free energy & 1 & 642 & 100.0\% & \cite{mobley2014freesolv}  \\
            & Lipophilicity  & $\mathrm{log} \ K_{\mathrm{octanol} - \mathrm{water}}$  & 1 & 4,194 & 99.9\% & \cite{mendez2019chembl, lipo} \\
            \midrule
            Biophysics & BACE & BACE-1 inhibition & 1 & 1,511 & 99.9\% & \cite{subramanian2016computational} \\
            \midrule
            & BBBP & Blood-brain barrier penetration & 1 & 1,959 & 99.2\% & \cite{martins2012bayesian}  \\
            & Tox21 & Qualitative toxicity & 12 & 7,677 & 98.0\% &  \cite{tox21} \\
            Physiology & ToxCast  & Qualitative toxicity & 617 & 8,405 & 98.0\% & \cite{richard2016toxcast}  \\
            & SIDER & Drug side effects & 27 & 1,356 & 95.1\% & \cite{kuhn2016sider} \\
            & ClinTox  & Toxicity of failed, approved drugs & 2 & 1,438 & 98.7\% & \cite{novick2013sweetlead, aact} \\
            \bottomrule
        \end{tabular}

	\caption{Experimental data for GEOM species from MoleculeNet \cite{Wu2018MoleculeNet}. ``Species'' denotes the number of MoleculeNet compounds that have CREST CREs in vacuum. ``Recovered'' gives this quantity as a percentage of the original number of compounds in MoleculeNet. The original numbers in each dataset, used to compute the ``recovered'' percentage, are slightly different than in Ref. \cite{Wu2018MoleculeNet}. This is because several of the original compounds were found to be identical after SMILES pre-processing and conversion to InChi keys. Note that 1,511 BACE species (99.9\%) also have CREST CREs in water.}
	\label{tab:molecule_net}
\end{table}

\setlength{\tabcolsep}{10pt}
 \begin{figure}[H]
    \centering

    	\begin{tabular}{lccc}
            \toprule
            & & AICures drug dataset \\ \midrule
             & Mean & Std. deviation & Maximum \\ 
            \midrule
            $S$ (cal/mol K) & 8.2  & 2.6 &  16.8 \\
            -$G$ (kcal/mol) & 2.4 & 0.8 & 5.0 \\
            $\langle E\rangle$ (kcal/mol) & 0.4 & 0.2  & 2.4 \\
            Conformers & 102.6 & 159.1 & 7,451 \\
            \midrule
            & & QM9 dataset \\ \midrule
             & Mean & Std. deviation & Maximum \\ \midrule
            $S$ (cal/mol K)  & 3.9  & 2.8 & 14.2 \\
            -$G$ (kcal/mol) & 1.2 & 0.8 & 4.2 \\
            $\langle E  \rangle$ (kcal/mol) & 0.2 & 0.2  & 2.2 \\
            Conformers & 13.5 & 42.2 & 1,101 \\
            \bottomrule
        \end{tabular}

    \caption{CREST-based statistics for the QM9 and AICures drug datasets.}
    \label{tab:data_stats}

  \end{figure}
 
\begin{table}[H]
	\vspace{6mm}
    \centering
    	\begin{tabular}{lccc}
            \toprule
            Model & $G$ (kcal/mol)  & $\braket{E}$ (kcal/mol) & $\mathrm{ln}$(unique conformers)  \\  
            \midrule
            Random Forest &  0.406 & 0.166 &  0.763 \\
            KRR & 0.289 &  0.131 & 0.484  \\ 
            FFNN & 0.274  & 0.119 & 0.455\\ 
            ChemProp & 0.225 & \textbf{0.110} & 0.380  \\
            SchNetFeatures & \textbf{0.203} & 0.113 & \textbf{0.363} \\
            \bottomrule
        \end{tabular}
    \caption{Prediction mean absolute error (MAE) for three conformer-related properties. Models were trained and tested on the AICures drug dataset. }
    \label{tab:results}
\end{table}

\end{document}